\documentclass[prl,twocolumn,superscriptaddress]{revtex4}
\usepackage{}%
\usepackage{amsfonts}
\usepackage{amsmath}
\usepackage{amssymb}
\usepackage{graphicx}%
\usepackage{color}
\emergencystretch=\maxdimen
\begin{document}

\title{Phonon vibrational and transport properties of SnSe/SnS superlattice at finite temperatures}
\author{Feng-ning Xue}
\affiliation{College of Mathematics and Physics, Beijing University of Chemical Technology, Beijing 100029, People's Republic of China}
\affiliation{State Key Laboratory of Explosion Science and Technology, Beijing Institute of Technology, Beijing 100081, People's Republic of China}
\author{Wei Li}
\affiliation{School of Electronic Information Engineering, Lanzhou Institute of Technology, Lanzhou 730050, People's Republic of China}
\author{Zi Li}
\affiliation{Institute of Applied Physics and Computational Mathematics and National Key Laboratory of Computational Physics, Beijing 100088, People's Republic of China}
\author{Yong Lu}
\thanks{Corresponding author; luy@mail.buct.edu.cn}
\affiliation{College of Mathematics and Physics, Beijing University of Chemical Technology, Beijing 100029, People's Republic of China}
%\pacs{63.20.Ry, 63.20.kg, 63.20.dk}

\date{\today}% It is always \today, today,
%\pacs{61.50.-f, 74.10.+v, 63.20.dk}
%63.20.D- %Phonon states and bands, normal modes, and phonon dispersion
%63.20.Ry %Anharmonic lattice modes
%63.20.dk %First-principles theory
%63.20.kg %Phonon-phonon interactions}
\clearpage

\begin{abstract}
The structural stability and phonon properties of SnSe/SnS superlattices at finite temperatures have been studied using machine learning force field molecular dynamics and the anharmonic phonon approach. The vertical SnSe/SnS superlattice undergoes a phase transition from the Pnma phase to a novel P4/nmm phase at finite temperatures, which is different from the high-temperature Cmcm phase of the SnSe and SnS systems. The stability of P4/nmm phase is determined by molecular dynamics trajectories and anharmonic phonon dispersion relations. The imaginary modes of TO modes at the $\textbf{q}$=M($\frac{1}{2},\frac{1}{2},0$) point of the P4/nmm phase in harmonic approximation become rigid at elevated temperatures. An analysis of phonon power spectra upon temperature also confirms the dynamic stabilization. The P4/nmm phase has higher symmetry than the Pnma phase, and the phase transition between them is accompanied by competition between the Jahn-Teller effect and phonon anharmonicity. Unlike the anisotropic distribution of Sn-Se/S bonds in the Pnma phase, the P4/nmm phase forms chemical bonds with similar bond lengths both in-plane and interlayer, and their resonance effect can significantly enhance phonon scattering. The calculated phonon density of states and lifetime is strongly temperature dependent, demonstrating the heavy anharmonicity in the SnSe/SnS system. The P4/nmm phase has an extremely low lattice thermal conductivity, close to the experimental values of SnSe and SnS. Moreover, with the reduction of band gap and the enhancement of band degeneracy near the Fermi level, the P4/nmm phase exhibits superior electronic transport properties and significantly enhanced response to infrared and visible light. This makes it show great potential in thermoelectric and photovoltaic applications.
\end{abstract}

\maketitle
\clearpage
\section{I. INTRODUCTION}
Layered IV-VI group semiconductors have attracted widespread attention in thermoelectric and optoelectronic fields in recent years due to their excellent physical transport properties \cite{Zhao2024,Agata2024,Liu2022,Sutter2021,Xie2021,Kai2022,Michel2022,Ya2023}. The strong in-plane covalent bonding and weak interlayer van der Waals (vdW) interactions allow for the easy formation of low-dimensional materials and heterostructures with high-quality interfaces and nanoscale thickness \cite{Li2022,Liu2023,Mao2024}. As an important member of this family, SnSe and SnS have garnered particular interest due to their superior thermoelectric performance and optoelectronic capability \cite{Raturi2024,LinN2024,Lo2024,Zhang2023}. SnSe and SnS, sharing the same crystal structure, undergo a structural phase transition from the Pnma phase to the Cmcm phase at high temperatures. Their intrinsic low lattice thermal conductivity and the strong phonon interactions associated with the phase transition make them an ideal platform for studying the fundamental principles of phonon anharmonicity and transport properties \cite{Li2015,Lu2019,Aseginolaza2019,Lanigan2020}.

Currently, there have been numerous reports on improving the transport properties of SnSe and SnS by defect engineering and doping, alloying, and nanostructure design \cite{Yang2023,Sheng2023,Damla2023,Zhang2024,LuH2024}. Especially, SnSe and SnS have compatible lattice constants \cite{Zhao2014,Chen2003}, allowing them to be well-connected through interlayer vdW interactions to construct nanoscale heterostructures and superlattices. This enables the optimization of their electronic and phonon transport properties by precisely controlling the microstructure of the SnSe/SnS heterosystems. Experimentally, various superlattice structures based on SnSe or SnS have been successfully fabricated, demonstrating the experimental feasibility of this regulatory method \cite{Robert2020,Zhao2022,Gao2022,Schulz2024}. For example, Roberts \emph{et al.} \cite{Robert2020} designed amorphous precursor thin films to precisely control the structure and composition of SnS/TaS$_2$ superlattices; Zhao \emph{et al.} \cite{Zhao2022} reported an ordered CoSe$_2$/SnSe superlattice synthesized by vdW space-limited epitaxial growth method, which exhibits extremely low in-plane thermal conductivity and optimized thermoelectric performance; Gao \emph{et al.} \cite{Gao2022} successfully prepared $n$-type PbSe/SnSe superlattice structures by pulsed laser deposition technology, showing ultra-high power factors and extremely low lattice thermal conductivity at room temperature. Most recently, Schulz and colleagues \cite{Schulz2024} proposed a novel, simple, and reproducible one-pot heating technique for synthesizing two-dimensional SnSe/SnS core/crown heterostructured nanosheets. The advancement of these fabrication techniques provides a practical basis for the theoretical design and study of SnSe/SnS superlattices.

At present, theoretical research on the SnSe/SnS system has primarily focused on bilayer heterostructures \cite{Low2018,Li2020,Raturi2024}. For instance, Low \emph{et al.} \cite{Low2018} constructed different types of bilayer SnSe/SnS heterostructures using first-principles methods, which show potential optical absorption characteristics in the infrared region; Li \emph{et al.} \cite{Li2020} studied the electronic structure and thermoelectric properties of bilayer SnSe/SnS heterostructures, revealing the significant impact of interlayer interactions on SnSe/SnS heterostructures. These studies provide insights into the physical properties of low-temperature SnSe/SnS heterojunctions. According to our previous researches, temperature and the number of atomic layers can significantly affect the phase stability and transport properties of the SnSe system \cite{Lu2019,Lu2021,Xue2022}. The competition between phonon anharmonic effects and Jahn-Teller effects leads to significant differences in phase stability and transport properties for bulk, few-layer, and monolayer SnSe. Therefore, studying the dynamic stability and phase transition behavior of SnSe/SnS superlattice at finite temperatures is crucial for understanding their transport properties. Temperature effect is also an environmental factor that must be considered for the device applications.

In this study, we constructed vertical SnSe/SnS superlattice structures based on the low-temperature Pnma and high-temperature Cmcm phases as initial structures. The temperature effects on their dynamic stability and transport properties were investigated through first-principles calculations and machine learning force field molecular dynamics (FFMD) simulations. Surprisingly, unlike the high-temperature phase transition behavior of the SnSe and SnS systems, the SnSe/SnS superlattice structures do not transition to the Cmcm phase at finite temperatures but instead form a novel P4/nmm phase. This phase is significantly different from the Pnma phase in terms of atomic and electronic structures. By integrating large-scale FFMD simulations with the anharmonic phonon approach, we systematically investigated the phonon vibrational and transport properties of this phase. Additionally, we calculated the electronic structures and thermoelectric properties to assess their potential in thermoelectric and optoelectronic applications.

\section{II. Methods and Computational Details}
The SnSe/SnS superlattice structures were constructed using the Pnma and Cmcm phases as initial structures along the out-of-plane direction. Structural optimization and electronic structure calculations were performed using density functional theory (DFT) with the projector-augmented-wave (PAW) method \cite{Blochl1994}, as implemented in the VASP software package \cite{Kresse1999}. The electron exchange and correlation potential was modeled using the generalized gradient approximation (GGA) with parameters from Perdew, Burke, and Ernzerhof (PBE) \cite{PBE1996}. A Monkhorst-Pack 12$\times$12$\times$3 $k$-point mesh was employed for Brillouin zone (BZ) integration \cite{Monkhorst1976}. The plane-wave cutoff energy was set at 450 eV, ensuring energy convergence to within 10$^{-5}$ eV/atom. For the computation of harmonic second-order interatomic force constants, 4$\times$4$\times$2 supercells were used, where the finite displacement method with a displacement amplitude of 0.01 \AA$~$was utilized. The harmonic phonon frequencies and polarization vectors were derived using the PHONOPY post-processing software \cite{Togo2010}.

To calculate the anharmonic phonon transport properties, we employed large-scale machine learning force field molecular dynamics simulations, based on Bayesian linear regression to train the force field potential \cite{JinnouchiPRL2019,JinnouchiPRB2019,Jinnouchi2020}. The dataset for the machine learning force field was created through ab-initio molecular dynamics (AIMD) simulations using the VASP (version 6.3.0) package. The cutoff radii for radial and angular descriptors were set at 8.0 \AA$~$and 5.0 \AA, respectively. For dataset construction, the primitive unit cell was expanded to a supercell with dimensions 4$\times$4$\times$2, containing 256 atoms. In MD simulations, the package collects data in real-time and gradually builds a training set. When the errors in the energy and force predicted by the current force field exceed the preset threshold, the program performs a new AIMD calculation and adds this new reference data to the training set. This method ensures the accuracy of the force field and avoids unnecessary DFT calculations in most cases, thereby significantly accelerating the simulation process. A Bayesian approach is used to estimate the error of the current force field. After each MD step, the system dynamically adjusts the error threshold based on the average error of the previous 10 time steps. This adjustment mechanism is particularly important in high-temperature environments because, at high temperatures, the absolute values of forces are larger, which may lead to more configurations being sampled. The machine learning force field is constructed by dynamically expanding the local reference configurations. As the simulation progresses, new local reference configurations are added to the base set of the force field. When enough new reference data has been accumulated, the force field is updated to improve its predictive ability in subsequent simulations.

To enhance the efficiency of machine learning sampling, the training began with simulations at 800 K within the canonical ensemble (NVT) at the equilibrium volume, followed by further AIMD sampling at 5\% volume expansion and compression. Each simulation ran for 60 ps with a time step of 2 fs. Subsequently, a series of AIMD simulations were performed, annealing from 700 to 300 K, with each lasting 20 ps. A 2$\times$2$\times$1 $k$-point mesh centered at the $\Gamma$ point was used for all simulations. After training, the Bayesian error estimates for energy (BEEE) and force (BEEF) converged to 0.5$\times$10$^{-6}$ eV/atom and 5$\times$10$^{-3}$ eV/\AA, respectively, as shown in the Supplemental Material Fig. S1 \cite{Supplement}. To verify the accuracy of the force field, comparisons were made with AIMD results. As depicted in the Supplemental Material Fig. S2 \cite{Supplement}, the pair correlation function $g$($r$) at 300 K shows that the functions from both FFMD and AIMD simulations are in good agreement. The Supplemental Material Fig. S3 \cite{Supplement} shows the consistent atomic displacement probability distributions of Sn, Se, and S atoms calculated by FFMD and AIMD simulations at 300 K. All the Sn, Se and S atoms exhibit vibrations around the equilibrium lattice points, displaying a Gaussian function distribution, thereby corroborating the precision of the force field functions.

Temperature-dependent phonon frequencies, power spectra, lifetimes, and lattice thermal conductivity were derived using the anharmonic phonon approach based on the framework of machine learning FFMD and lattice dynamics \cite{Lu2024}. This approach enables the extraction of anharmonic phonon frequencies and lifetimes from velocity auto-correlation functions (VAF) in reciprocal ($q$) space that is defined as
\begin{equation}
\langle v^{*}_{\lambda}(0) v_{\lambda}(t)\rangle=\lim\limits_{t_{0}\rightarrow{\infty}} \frac{1}{t_0} \int^{t_{0}}_{0} v_{\lambda}(t')v_{\lambda}(t'+t)dt',
\end{equation}
where the mode-projected velocity, $v_{\lambda}(t)$, can be expressed as
\begin{equation}
v_{\lambda}(t)=\sum_{j=1}^{N}\sqrt{M_j} v_{j}(t)e^{-i\textbf{q}\cdot \textbf{R}_{j}}\cdot \hat{{e}}_{\lambda}.
\end{equation}
Here, $v_{j}(t)$ ($j=1,...,N$) are atomic velocities obtained from MD with $N$ atoms per supercell. $\hat{{e}}_{\lambda}$ is the polarization vector of the phonon mode $\lambda$. $M_j$ and $\textbf{R}_j$ are the atomic mass and coordinates of the $j$th atom in the supercell. Due to the anharmonic effect, the VAF shows the damped harmonic oscillator with an associated exponential function describing the mode lifetime, which can be well fitted by the equation
\begin{equation}
\langle v^{*}_{\lambda}(0) v_{\lambda}(t)\rangle=k_{B}Tcos(\widetilde{\omega}_{\lambda}t)e^{-\Gamma_{\lambda}t},
\end{equation}
where the anharmonic phonon frequency $\widetilde{\omega}_{\lambda}$ and the phonon linewidth $\Gamma_{\lambda}$ of all individual phonon modes can be extracted. To balance convergence and efficiency, we conducted FFMD simulations on 12$\times$12$\times$2 and 4$\times$4$\times$12 supercells to obtain the anharmonic phonon properties along the $xy$-plane and $z$-axis directions, respectively. We employed the Nos$\acute{e}$-Hoover thermostat to control temperature fluctuations \cite{Nose1984}. Each FFMD simulation lasted 60 ps with a time step of 2 fs. The volumetric thermal expansion at a specific temperature was determined by calculating the energy-volume curve through a series of FFMD simulations. The lattice thermal conductivity tensor \( \kappa^{\alpha \beta} \) was calculated according to \cite{Lu2024}
\begin{eqnarray}
\kappa^{\alpha \beta}=\sum_{\lambda} C_v(\lambda)  v^{\alpha}_{\lambda} v^{\beta}_{\lambda} \tau_{\lambda},
\end{eqnarray}
where $v^{\alpha(\beta)}_{\lambda}$ is the group velocity component along the $\alpha(\beta)$-direction; $\tau_{\lambda}$ is the phonon lifetime of mode $\lambda$; $C_v(\lambda)$ is the mode specific capacity of the following expression,
\begin{eqnarray}
C_v(\lambda)=\frac{k_\textup{B}}{N_{q} \Omega} (\frac{\hbar \omega_{\lambda}}{k_B T})^{2} f_0 (f_0 + 1),
\end{eqnarray}
where $k_\textup{B}$ is the Boltzmann constant, $N_q$ is the number of $q$ points in the Brillouin zone, $\Omega$ is the volume of the unit cell, and $f_0$ is the equilibrium Bose-Einstein distribution function.

\begin{figure*}
\centering
\includegraphics[width=1.4\columnwidth]{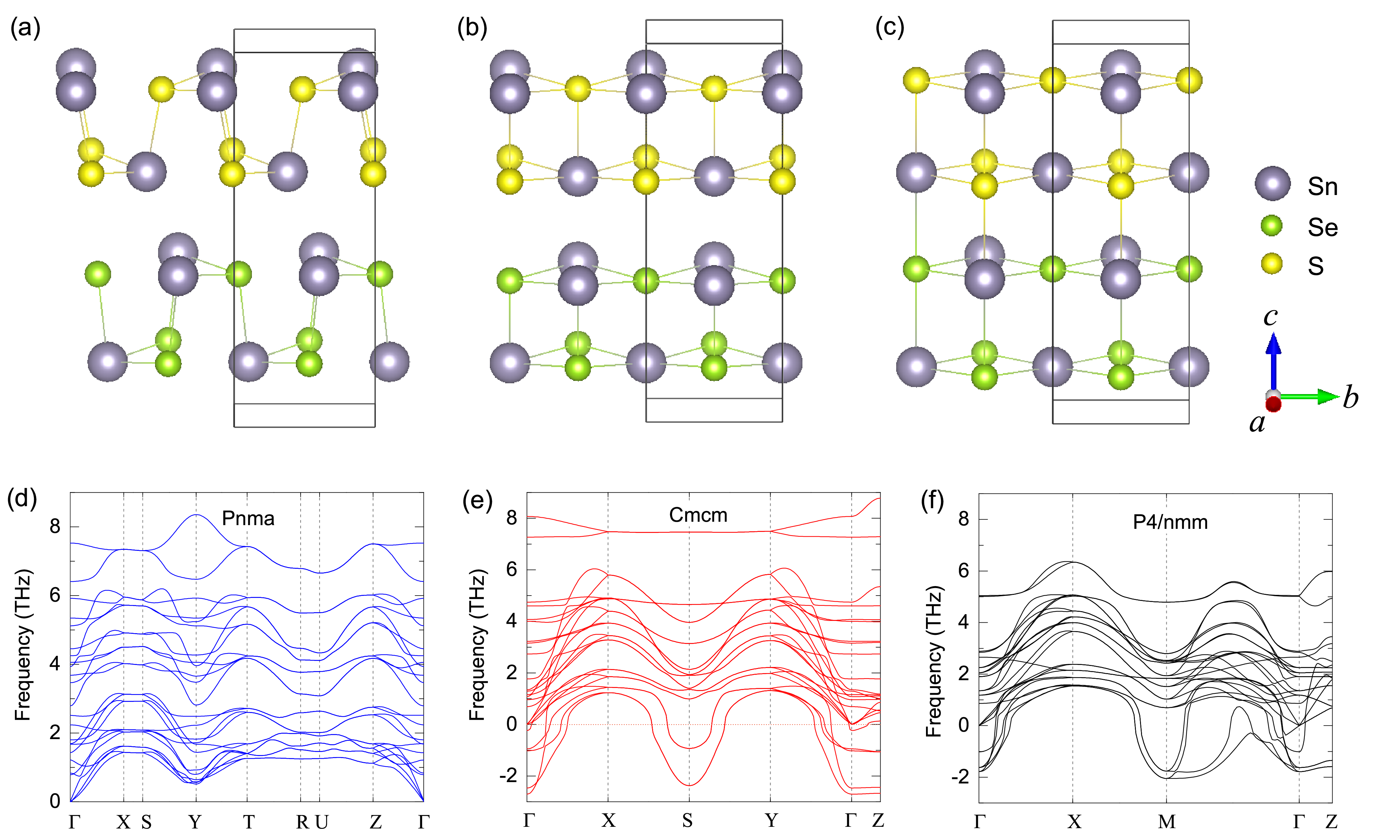}
\caption{(Color online) The crystal structures of SnSe/SnS superlattice in (a) Pnma, (b) Cmcm, and (c) P4/nmm phases. The same crystal axes are chosen to facilitate the comparison between these phases. (d)-(f) show the corresponding harmonic phonon dispersion relations along high-symmetry BZ directions of the three phases.
\label{fig:Graph1}}
\end{figure*}

\section{III. Results and discussions}
\subsection{A. Dynamic stability at finite temperatures}

\begin{figure*}
\includegraphics[width=1.4\columnwidth]{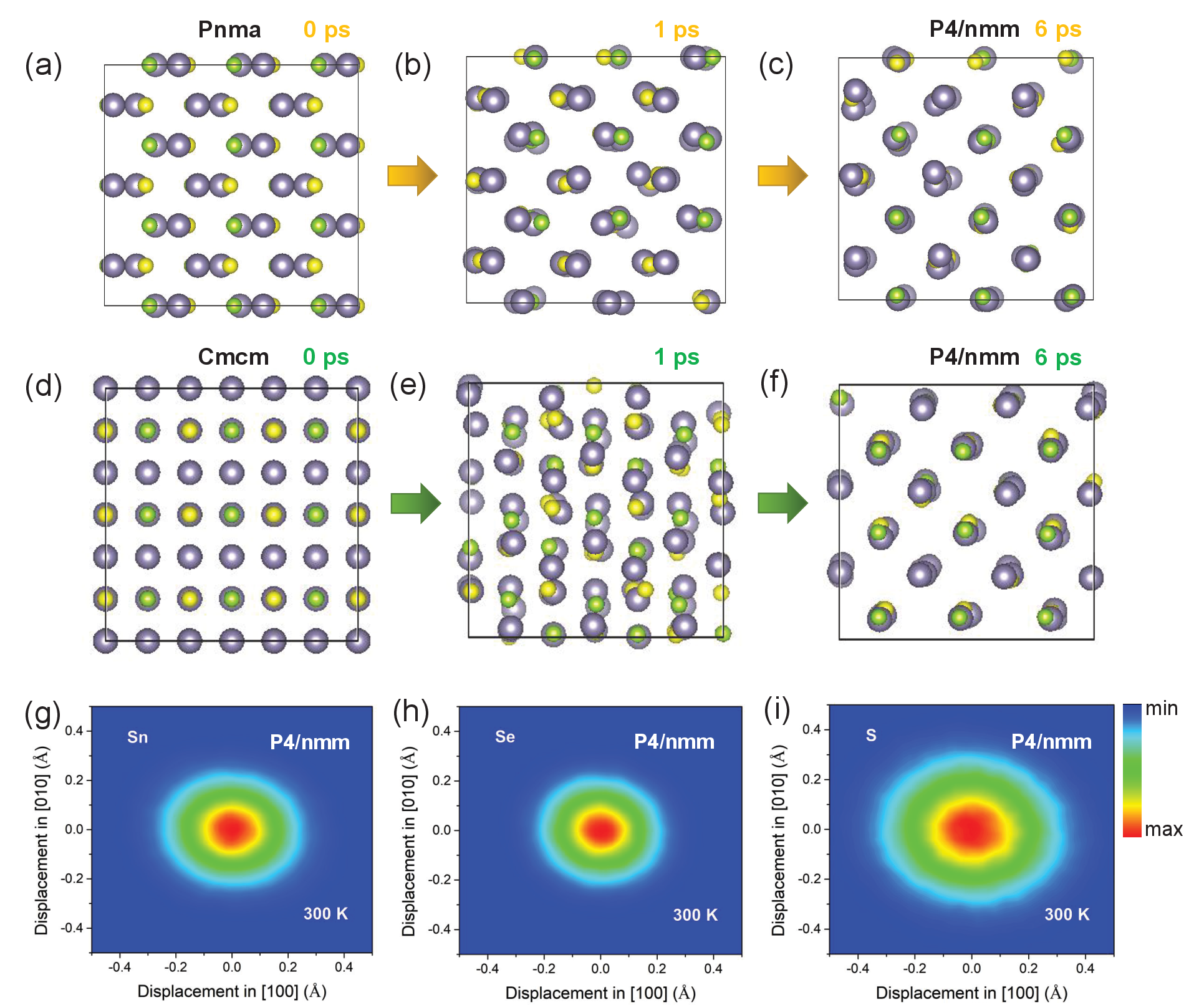}
\caption{(Color online) The atomic-scale evolutions of SnSe/SnS superlattices are shown for the (a)-(c) Pnma and (d)-(f) Cmcm phases, starting from the initial structure at 300 K. Snapshots from AIMD simulations along the $z$-axis at 0, 1, and 6 ps are displayed in (a)-(c) and (d)-(f), respectively. Probability distributions of atomic displacements projected onto the $xy$-plane, relative to the ideal P4/nmm structure, are presented for (g) Sn, (h) Se, and (i) S atoms at 300 K. }
\end{figure*}

\begin{figure*}
\centering
\includegraphics[width=1.4\columnwidth]{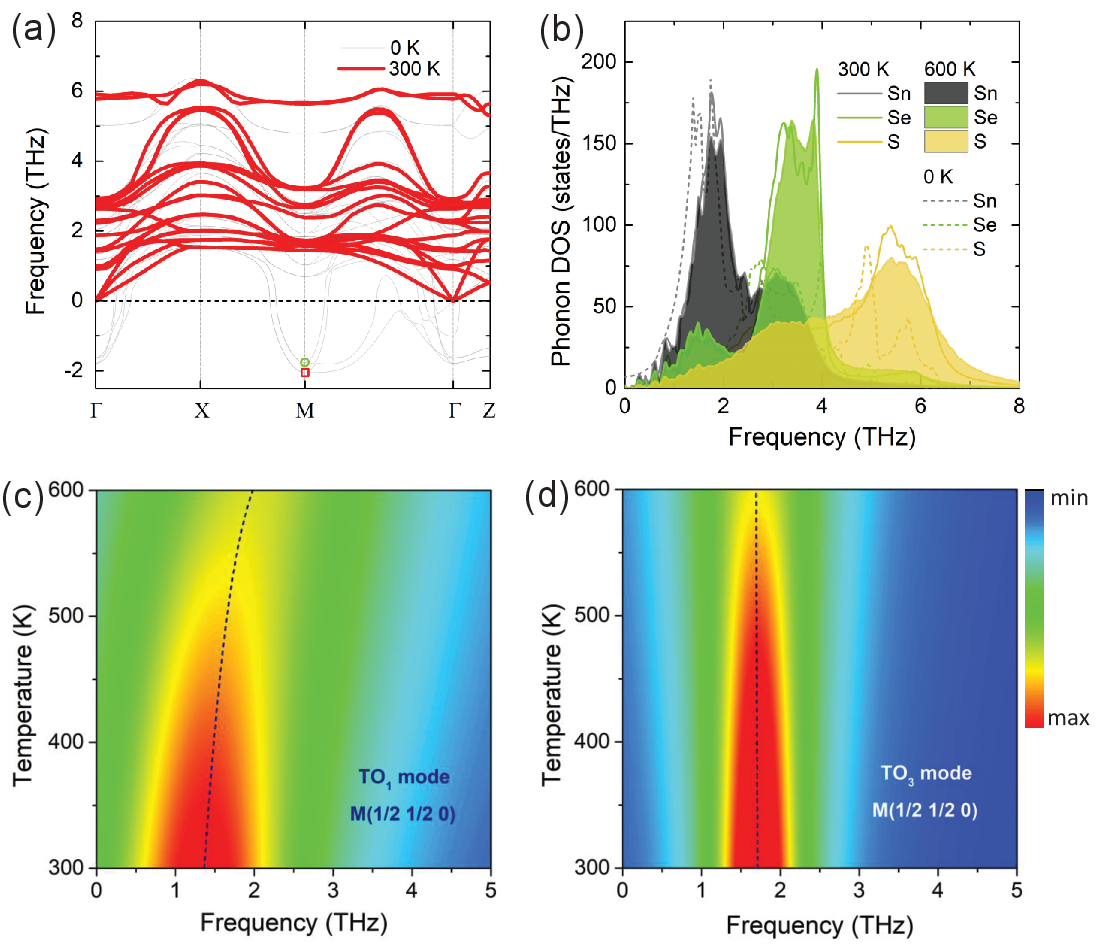}
\caption{(Color online) (a) Calculated anharmonic phonon dispersion curves for the P4/nmm phase at 300 K, with harmonic results at 0 K shown for comparison. (b) Calculated phonon DOS obtained by Fourier transforming the VAF from FFMD simulations at 300 and 600 K, with harmonic results at 0 K shown for comparison. (c) and (d) Calculated phonon power spectra of TO$_1$ and TO$_3$ modes at the $\textbf{q}$=M(1/2 1/2 0) point, marked by the red square and blue circle in (a), respectively, as a function of temperature. The color bar in (d) indicates the linewidth values for these two modes.
\label{fig:Graph1}}
\end{figure*}

The crystal structure of the SnSe/SnS superlattice for the low-temperature Pnma phase and the high-temperature Cmcm phase are illustrated in Fig. 1(a) and (b), respectively. The same crystal axes are chosen to facilitate comparison between these two phases. The optimized lattice parameters are $a$=4.17 \AA, $b$=4.39 \AA, and $c$=11.83 \AA$~$for the Pnma phase and $a$=4.21 \AA, $b$=4.22 \AA, and $c$=11.88 \AA$~$for the Cmcm phase, respectively. The Pnma results are close to the experimental results of SnSe and SnS, with SnSe having $a$=4.22 \AA, $b$=4.40 \AA, and $c$=11.58 \AA$~$\cite{Zhao2014}, and SnS having $a$=3.99 \AA, $b$=4.33 \AA, and $c$=11.20 \AA$~$\cite{Chen2003}.  The symmetry of the Cmcm phase is higher than that of the Pnma phase, mainly attributed to the distortion of Sn atoms along the $b$ direction in the Pnma phase. The harmonic phonon dispersion of the Pnma and Cmcm phases was calculated and shown in Fig. 1(d) and (e). The phonon dispersion relations of the low-temperature Pnma phase are stable throughout the BZ. In contrast, imaginary phonon frequencies appear evidently in the high-symmetry Cmcm phase. The Pnma phase is the preferred form of the SnSe/SnS superlattice at low temperatures, in agreement with that of the SnSe and SnS systems.

To confirm the stability of these phases at finite temperatures, we performed AIMD simulations of these two phases at 300 K. As shown in Fig. 2(a) and (d), we set the Pnma and Cmcm phases as the initial structures, respectively. During the simulation process, both Pnma and Cmcm phases could not maintain their initial symmetry. The SnSe and SnS layers tend to migrate along the $b$-axis, as shown by the snapshot of MD simulations in Fig. 2(b) and (e). As a result, both phases transform into a more high-symmetry phase, as shown in Fig. 2(c) and (f). This novel phase belongs to the tetragonal P4/nmm phase with lattice parameters $a$=$b$=4.21 \AA$~$and $c$=11.89 \AA, consistent with the results of the Cmcm phase. As shown in Fig. 1(c), Sn atoms form octahedral structures with adjacent Se/S atoms in P4/nmm phase, different from the Pnma and Cmcm phases, which form tetrahedral structures. The typical layer gap between the SnSe and SnS layers is replaced by Sn-Se and Sn-S bonds. Unlike the SnSe and SnS bulk phases, which have the Cmcm phase as the high-temperature stable phase, the P4/nmm phase is the stable phase of the SnSe/SnS superlattice at finite temperatures. It is noted that the P4/nmm phase is a temperature-driven phase, since its harmonic phonon dispersion curves are unstable at low temperatures, as shown in Fig. 1(f). To confirm the structural stability of the P4/nmm phase with temperature, we performed FFMD simulations at finite temperatures. Based on the evolution relationship of atomic displacement over time obtained from FFMD simulations, we obtained the displacement probability distribution functions of Sn, Se, and S atoms by statistically analyzing the displacement distributions of all corresponding atoms in the simulation supercell. As shown in Fig. 2(g)-(i), all atoms were found to oscillate around their equilibrium positions with respect to the P4/nmm symmetry, with the root mean square displacement (RMSD) of 0.33 \AA$~$for Sn, 0.31 \AA$~$for Se, and 0.37 \AA$~$for S atoms at 300 K. These observations confirm the structural stability of the P4/nmm phase at room temperature. Upon further increasing the temperature to 600 K, the displacement amplitude significantly increases, with the RMSD increasing to 0.45 \AA$~$for Sn, 0.42 \AA$~$for Se, and 0.48 \AA$~$for S atoms, respectively. The phase stability is not affected.

\subsection{B. Anharmonic Phonon Vibrational Properties}
To obtain the anharmonic phonon dispersion relations at finite temperatures, we employed large-scale FFMD simulations to incorporate the temperature effect. The anharmonic phonon frequency was derived by fitting the \( q \)-space projected VCF in accordance with Equ. (1). Utilizing the renormalized phonon frequencies \( \widetilde{\omega}_{\mathbf{q},s} \) extracted from all the \( \mathbf{q} \) points sampled in the FFMD simulations, we determined the anharmonic phonon dispersion relations at 300 K, as depicted in Fig. 3(a). Taking into account the anharmonic interactions, the high-temperature phonon dispersion curves of the P4/nmm phase become stable, with the imaginary frequencies observed in harmonic calculations disappearing. To illustrate the temperature impact on phonon properties, including frequency and linewidth, we calculated the phonon power spectrum using the following equation,
\begin{equation}
G_{\lambda}(\omega) = \int_{0}^{\infty} \langle v^{*}(0)v(t)\rangle_{\lambda} \text{exp}(i\omega t)dt.
\end{equation}
We selected two typical phonon modes that exhibit significant imaginary frequencies at the \( \mathbf{q} \)=M(1/2 1/2 0) point with temperature, marked by the red square for TO$_1$ and the blue circle for TO$_3$ mode in Fig. 3(a), respectively. As shown in Fig. 3(c), the frequency of the TO$_1$ mode demonstrates a significant nonlinear increase from 1.35 THz to 1.98 THz with the temperature rising from 300 to 600 K, reflecting the anharmonic effect on frequency. The linewidth $\Gamma_{TO_1}$ also notably broadens from 2.25 THz to 3.96 THz, indicating enhanced phonon scattering at elevated temperatures. In contrast, the frequency of the TO$_3$ mode shows a weak temperature dependence, slightly increasing from 1.65 THz to 1.67 THz. However, the phonon linewidth for the TO$_3$ mode $\Gamma_{TO_3}$ also significantly broadens from 1.01 THz to 1.32 THz at higher temperatures, suggesting a reduced phonon lifetime at elevated temperatures.

The distribution characteristics of all phonon modes in \( q \)-space can be inferred from the phonon density of states (DOS). In Fig. 3(b), we present the phonon DOS of the P4/nmm phase at various temperatures, calculated by the Fourier transform of the VCF in real space obtained from the FFMD simulations. Three distinct peaks, distributed from low- to high-frequency and primarily contributed by Sn, Se, and S atoms, are clearly visible in the phonon DOS. This distribution characteristic is predominantly determined by the atomic masses of these elements. Nevertheless, a significant coupling between the Sn-Se and Sn-S bonds is observed near 1.5 and 3.0 THz, respectively, indicating the resonance of Sn-Se and Sn-S bonds. As the temperature increases, the main distribution characteristics remain unchanged, but the intensity of the distribution peaks diminishes. Compared to the harmonic results, the phonon DOS shifts significantly to higher frequencies, indicating the strong phonon anharmonicity. The full width at half maximum (FWHM) of the high-frequency peak is enlarged substantially. Due to the direct relationship between the FWHM and scattering rate, the high-frequency phonon exhibits more pronounced anharmonicity. It is noteworthy that the high-frequency peak is primarily contributed by the S atom. The strong anharmonicity is also evident in the vibrational amplitude of the S atoms, as shown in Fig. 2(i), which has the largest RMSD compared to the Sn and Se atoms.

\subsection{C. Phonon Transport Properties}
\begin{figure}
\includegraphics[width=0.8\columnwidth]{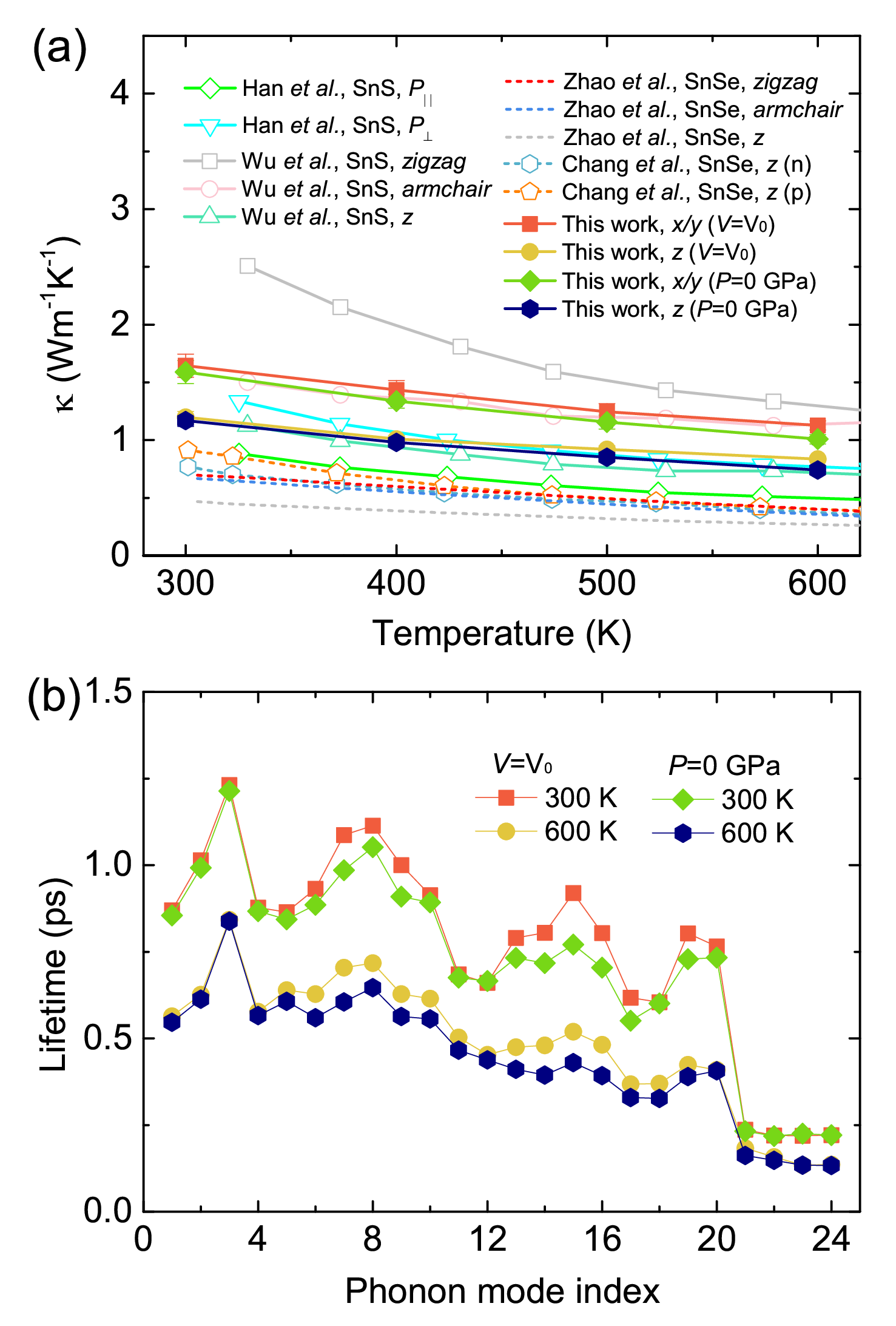}
\caption{(Color online) (a) Calculated lattice thermal conductivity $\kappa_{lat}$ (Wm$^{-1}$K$^{-1}$) of the P4/nmm phase along the $x/y$ and $z$ directions from 300 to 600 K. Experimental results for SnSe and SnS are included for comparison. (b) Predicted $q$-averaged phonon lifetime (ps) distribution for all 24 independent vibrational modes at 300 and 600 K under constant volume ($V$=$V_0$) and zero pressure ($P$=0 GPa).}
\end{figure}

Utilizing the Peierls-Boltzmann theory as outlined in Equ. (4), we have computed the lattice thermal conductivity (\(\kappa_\text{lat}\)) for the P4/nmm phase across a temperature span of 300 to 600 K. The temperature-dependent \(\kappa_\text{lat}\) is depicted in Fig. 4(a), with experimental data for SnSe and SnS included for comparison. At 300 K, the P4/nmm phase exhibits pronounced anisotropy in lattice thermal conductivity, with values of 1.65 Wm\(^{-1}\)K\(^{-1}\) in the $x/y$ direction and 1.19 Wm\(^{-1}\)K\(^{-1}\) along the $z$ direction under constant volume conditions.

Lattice thermal conductivity is intrinsically linked to phonon lifetimes. Fig. 4(b) illustrates the \(q\)-averaged phonon lifetime distribution for the 24 independent vibrational modes, indexed by their frequencies near the \(\Gamma\) point in ascending order. The phonon lifetimes across all modes vary from 0.2 to 1.3 ps. The impact of temperature on phonon lifetimes is pronounced. As the temperature rises from 300 to 600 K, phonon lifetimes diminish due to increased phonon scattering, leading to a decrease in \(\kappa_\text{lat}\) to 1.13 and 0.84 Wm\(^{-1}\)K\(^{-1}\) in the $x/y$ and $z$ directions, respectively.

Thermal expansion can slightly reduce phonon lifetimes and lattice thermal conductivity, as evidenced in Fig. 4. When accounting for volumetric expansion, the \(\kappa_\text{lat}\) values are modestly reduced, with a maximum decrease of 12\% at 600 K in the $x/y$ direction. Compared to the experimental results of SnS \cite{Han2015,Wu2018}, the \(\kappa_\text{lat}\) along the $x/y$ direction is similar to that of SnS along the armchair direction, while the $z$-direction \(\kappa_\text{lat}\) falls within the range of the out-of-plane ($z$) values of SnS . It is noteworthy that SnSe possesses inherently low thermal conductivity, a trait that significantly contributes to its exceptional thermoelectric performance. The \(\kappa_\text{lat}\) values of P4/nmm phase, ranging from 1.14 to 0.85 Wm\(^{-1}\)K\(^{-1}\), are close to the out-of-plane values of SnSe \cite{Zhao2014,Chang2018}, underscoring the potential of the P4/nmm phase for thermoelectric applications.

\subsection{D. Electronic Structure and Thermoelectronic Transport Properties}

\begin{figure}
\centering
\includegraphics[width=1.0\columnwidth]{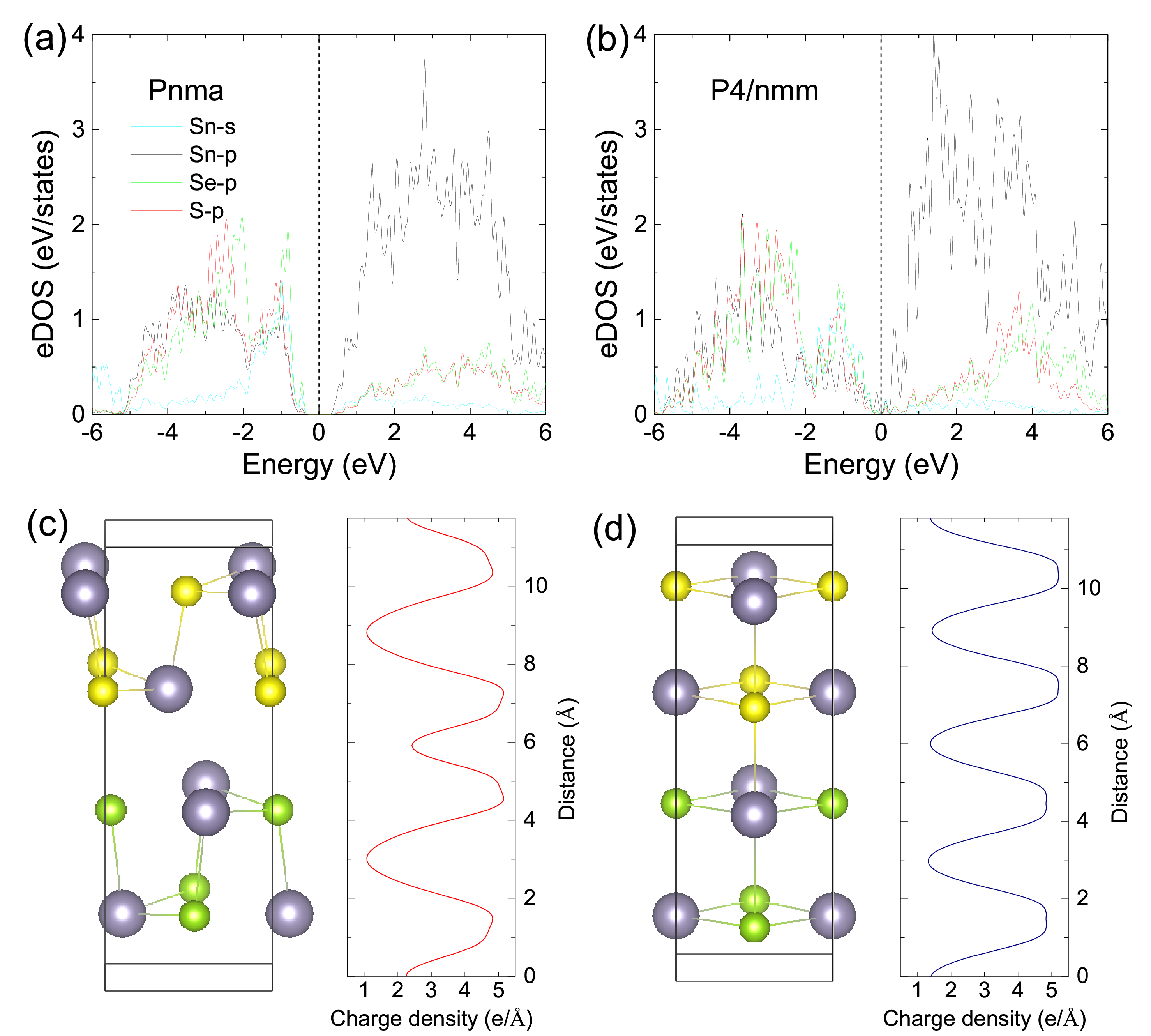}
\caption{(Color online) The electronic DOS for (a) Pnma phase and (b) P4/nmm phase, respectively. The linear distribution of electronic density along the $z$ bonding direction for (c) Pnma and (d) P4/nmm phases. The corresponding atomic structures are shown for comparison. \label{fig:Graph5}}
\end{figure}

\begin{figure}
\centering
\includegraphics[width=0.8\columnwidth]{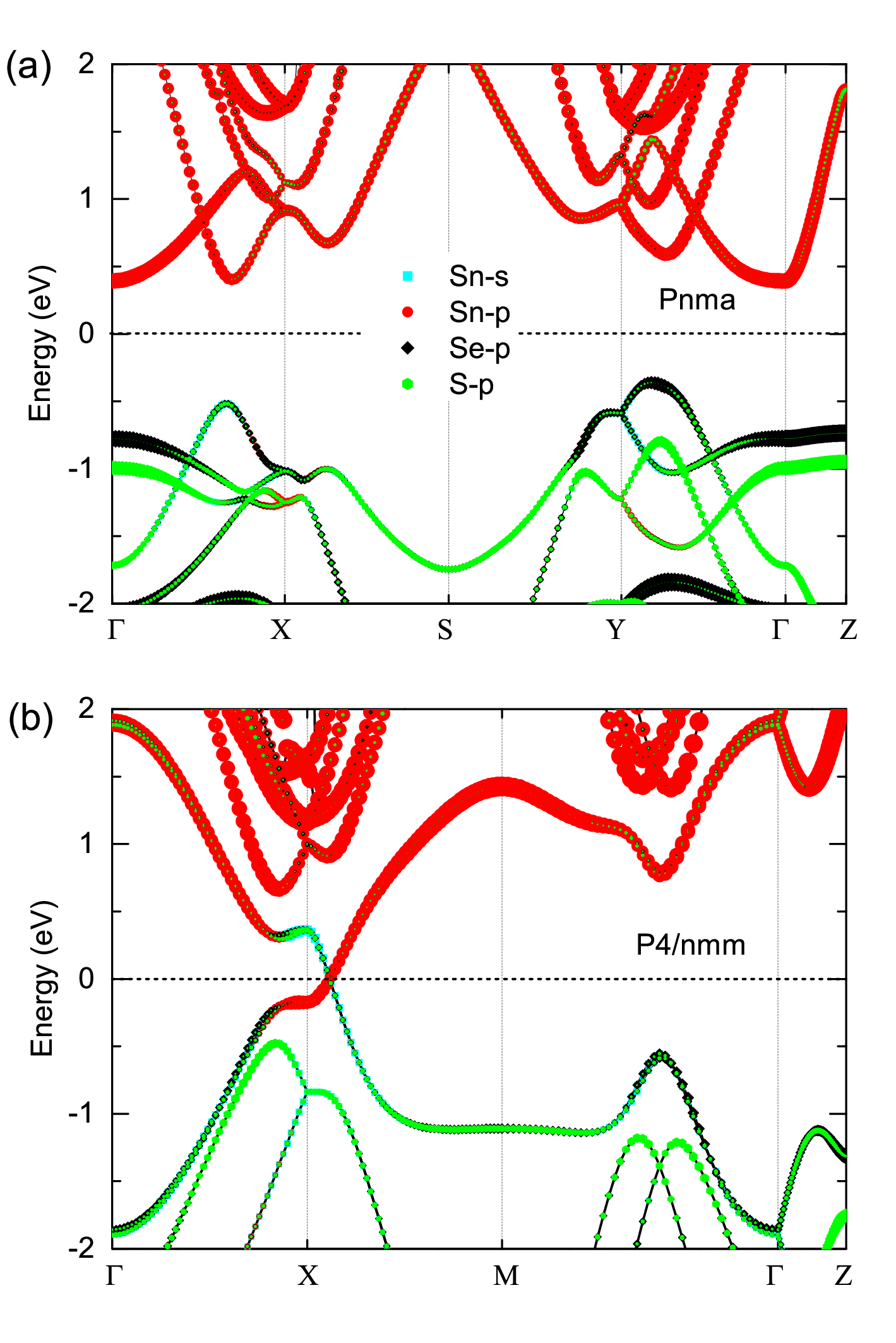}
\caption{(Color online) The orbital-projected band structures of (a) Pnma phase along the high-symmetry $\Gamma$-X-S-Y-$\Gamma$-Z path and (b) P4/nmm phase along the $\Gamma$-X-M-$\Gamma$-Z path in BZ respectively. The thickness of the curves hints the weight of the electronic orbital contributions.
\label{fig:Graph6}}
\end{figure}

\begin{figure*}
\centering
\includegraphics[width=1.4\columnwidth]{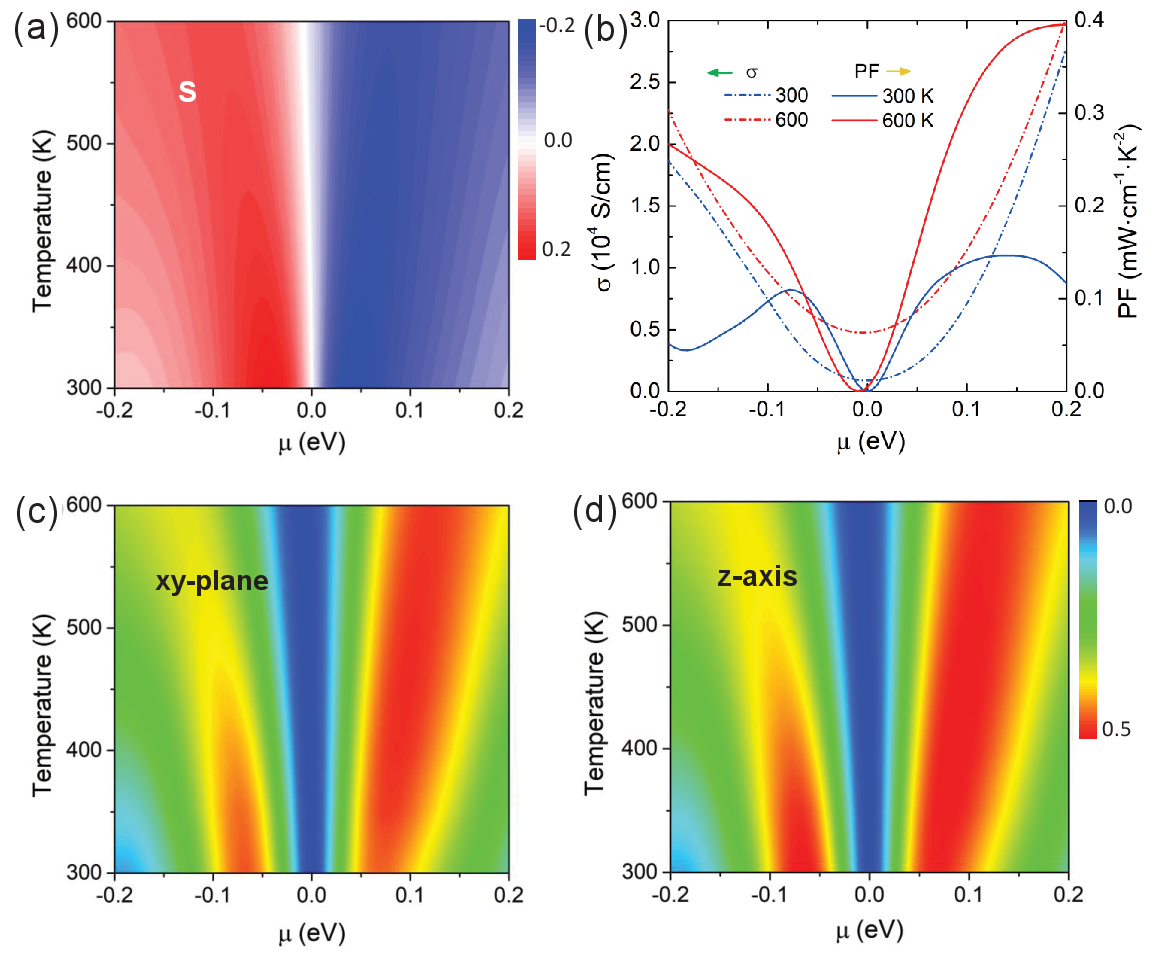}
\caption{(Color online) (a) The map of the Seebeck coefficient, $S$ (10$^{-3}$ V/K), as a function of temperature and chemical potential $\mu$ (eV). (b) Calculated electrical conductivity $\sigma$ (10$^4$ S/cm, left scale) and power factor PF (mWcm$^{-1}$K$^{-2}$, right scale) as a function of chemical potential at 300 and 600 K, respectively. (d) The ZT as a function of chemical potential. The negative and positive values of $\mu$ represent the $p$-type and $n$-type doping, respectively.
\label{fig:Graph7}}
\end{figure*}

As shown in Fig. 5(a), the orbital-projected electronic DOS indicates that the valence band of the Pnma phase is primarily composed of Se/S $p$-states hybridized with Sn $s$ and $p$-states, while the conduction band is mainly contributed by the Sn $p$-orbitals. The main characteristic peak of the valence band is located near -2.5 eV, with energy spread distributed across the range of -0.5 to -5.0 eV. A significant secondary localized peak forms near -1.0 eV at the valence band maximum (VBM). The main peak of the conduction band is located near 3.0 eV. A significant band gap of 0.7 eV is observed, indicating the semiconductor characteristics of the Pnma phase. Compared to the Pnma phase, the main peak of the valence band for the P4/nmm phase shifts to a lower energy position, near -3.0 eV as shown in Fig. 5(b). The distribution range of the valence band main peak extends to 0.0 to -6.0 eV, with reduced localization. The main peak position of the conduction band moves closer to the Fermi level, located near 1.8 eV. The band gap between the valence and conduction bands is occupied by electronic states, exhibiting weak metallic behavior. Bader effective charge analysis also shows that the ionic character of the Pnma phase is much higher than that of the P4/nmm phase. Specifically, in the Pnma phase, each Sn atom in the SnSe (SnS) atomic layer provides 2.13 (2.11) electrons to the surrounding Se/S atoms, while in the P4/nmm phase, it provides 1.05 (1.13) electrons.

Since electronic structure is closely related to light absorption characteristics, these significant changes in the distribution of characteristic peaks lead to differences in the light response capabilities of the two phases. To this end, we calculated and compared the optical absorption spectra of the Pnma and P4/nmm phases. As shown in Supplementary Material Fig. S4\cite{Supplement}, the spectral response of Pnma phase mainly starts from visible light, with the main absorption peaks distributed in the ultraviolet region of $\sim$5.0--7.0 eV, corresponding to the electron transition from the main characteristic peak of the valence band to that of the conduction band. In contrast, the P4/nmm phase has a significant absorption peak at 0.5 eV, showing a response to infrared light. Moreover, the main absorption peak of the P4/nmm phase is distributed in the violet and near-ultraviolet region of $\sim$3--5 eV, with the optical absorption spectrum intensity in the visible light range being nearly doubled compared to the Pnma phase. These significantly enhanced light response capabilities make the P4/nmm phase SnSe/SnS system highly promising for applications in solar cells, photocatalysis, and photodetectors.

As shown in Fig. 5(c), due to the lower symmetry of the Pnma phase structure, within the $xy$ plane, each Sn atom in the SnSe (SnS) atomic layer forms two shorter (stronger) bonds of 2.832 \AA$~$(2.724 \AA) and two longer (weaker) bonds of 3.270 \AA$~$(3.397 \AA) with the surrounding four Se (S) atoms. In the $z$-direction, the Sn atom only forms a chemical bond with one neighboring Se (S) atom with a bond length of 2.788 \AA$~$(2.639 \AA), while deviating from the $z$-direction on the other side to form a lone pair. Due to this asymmetric bonding distribution, there is a high charge density distribution between the SnSe and SnS atomic layers, as shown in Fig. 5(c). In contrast, in the P4/nmm phase, each Sn atom in the SnSe (SnS) atomic layer forms equivalent chemical bonds with the surrounding four Se (S) atoms, with bond lengths of 2.977 \AA$~$(2.976 \AA), as shown in Fig. 5(d). Moreover, in the $z$-direction, Sn forms bonds with the nearest Se and S on both sides, with bond lengths of 3.065 \AA$~$(3.018 \AA) and 2.917 \AA$~$(2.886 \AA), respectively. It is worth noting that the bond length differences of these six chemical bonds are very small, effectively forming bond resonance. Compared to the Pnma phase, the charge density distribution along the $z$-direction in the P4/nmm phase is more continuous, as shown in Fig. 5(d).

These differences in symmetry are also reflected in the changes of the corresponding electronic states. As shown in the orbital-projected band structure in Fig. 6, the bands of the P4/nmm phase are doubly degenerate near the VBM, while the degeneracy is lifted in the Pnma phase. According to the Jahn-Teller effect, structures with spatially degenerate electronic ground states undergo geometric distortion to eliminate degeneracy. Since distortion reduces the overall energy of the system, the Pnma phase is more stable than the P4/nmm phase at low temperatures. At high temperatures, the temperature effect can counteract the influence of geometric distortion through atomic thermal vibrations, thus promoting the stability of the high-symmetry phase. We also note that the band structure of the P4/nmm phase exhibits typical valley distribution characteristics along high-symmetry directions. Typically, high band degeneracy and multi-valley distribution in the valence band can increase the DOS near the Fermi level, which helps to enhance the Seebeck coefficient and electrical conductivity, thereby improving thermoelectric performance \cite{Zhao2014,Chang2018}. Therefore, we have also evaluated the thermoelectric transport properties, including the Seebeck coefficient ($S$), the electrical conductivity ($\sigma$), the power factor (PF=S$^2$$\sigma$), and the figure of merit (ZT), based on the Boltzmann semiclassical transport theory as implemented in the BOLTZTRAP code \cite{boltzmann2006}. The variation of there parameters with temperature and chemical potential is shown in Fig. 7(a)-(d). The PF can reach a maximum of 0.11 and 0.15 mWcm$^{-1}$K$^{-2}$ at 300 K for $p$-type and $n$-type, respectively. These values corresponds to a small chemical potential shift of -0.08 and 0.12 eV respectively, indicating that the transport properties of P4/nmm phase are sensitive to carrier concentration. At 300 K, along the $xy$ direction, the maximum ZT values for $p$-type and $n$-type are 0.42 and 0.46, respectively. Due to the lower lattice thermal conductivity in the $z$ direction, the maximum ZT values for $p$-type and $n$-type are significantly enhanced, with values of 0.46 and 0.51, respectively. Moreover, as the temperature increases to 600 K, $n$-type doping can still maintain a stable high ZT value above 0.50. These results indicate that the P4/nmm phase SnSe/SnS superlattice is an excellent room-temperature thermoelectric material.

\section{IV. Summary}
In this study, we utilized machine learning FFMD and the anharmonic phonon approach to investigate the structural stability and phonon properties of SnSe/SnS superlattices at finite temperatures. The research found that SnSe/SnS superlattices undergo a phase transition from the Pnma phase to a novel high-temperature P4/nmm phase at finite temperatures, distinct from the conventional high-temperature Cmcm phase of SnSe and SnS. This phase transition is accompanied by competition between the Jahn-Teller effect and phonon anharmonicity. The P4/nmm phase, which has higher symmetry than the Pnma phase, forms chemical bonds with similar bond lengths, and its resonance effect significantly enhances phonon scattering. The calculated vibrational mode power spectra and phonon DOS are strongly temperature-dependent, demonstrating strong anharmonicity in the SnSe/SnS system. The P4/nmm phase has an extremely low lattice thermal conductivity, with interlayer thermal conductivity values ranging from 1.14 to 0.85 Wm$^{-1}$K$^{-1}$ between 300 and 600 K, close to the experimental values of SnSe and SnS. Moreover, due to the enhancement of band degeneracy near the Fermi level brought about by the phase transition, the P4/nmm phase exhibits superior electronic transport properties. The $p$-type and $n$-type doping at room temperature achieve the highest thermoelectric ZT values of 0.46 and 0.51, respectively, demonstrating excellent room-temperature thermoelectric conversion performance. Additionally, due to the reduction in band gap values and the shift of the conduction band peak towards the Fermi level, the P4/nmm phase shows a significantly enhanced response to infrared and visible light, which also indicates its great potential for application in the photovoltaic field.

This work is supported by the National Natural Science Foundation of China under Grant No. 12074028 and the National Key R\&D Program of China under Grant No. 2022YFA1403103.

\end{document}